\documentclass[reprint,amsmath,amssymb,amsfonts,showpacs,superscriptaddress,aps,prl]{revtex4-1}
\usepackage{float}
\usepackage{graphicx}
\usepackage{subfigure}
\usepackage{dcolumn}
\usepackage{bm}
\usepackage[citecolor=blue, filecolor=blue, linkcolor=blue, urlcolor=blue, colorlinks, pdfhighlight=/I]{hyperref}
\begin{document}

\title{Exotic Multi-fold Vortex Lattices of Spin-Orbit Coupled Bose-Einstein Condensates in Optical Lattices}
\author{Ben Li}
\affiliation{%
Beijing National Laboratory for Condensed Matter Physics, Institute of Physics,
Chinese Academy of Sciences, Beijing 100190, China
}%
\author{Shu Chen}
\affiliation{%
Beijing National Laboratory for Condensed Matter Physics, Institute of Physics,
Chinese Academy of Sciences, Beijing 100190, China
}
\affiliation{School of Physical Sciences, University of Chinese Academy of Sciences, Beijing, 100049, China}
\affiliation{Collaborative Innovation Center of Quantum Matter, Beijing, China}
\date{\today}

\begin{abstract}
We investigate the ground state of two-dimensional Bose-Einstein condensates with Rashba spin-orbit coupling in square optical lattices and demonstrate the existence of rich phases with different lattice structures,
which is closely related to the degenerate structure of  single particle energy spectrum induced by the competition of spin-orbit coupling and optical lattices. 
We find that the ground state is in the phase with either parity-time-reversal or parity symmetry by direct numerical simulation. 
We show the phase diagram of ground state in the whole regime of spin-orbit coupling strength, and particularly find that the system supports multi-fold vortex lattices, in which ground state holds half-quantum vortex lattices, vortex-antivortex pair lattices and fundamental vortex lattices, simultaneously, when single particle energy minimums touch the boundary of the first Brillouin zone.

\begin{description}
\item[PACS numbers]
03.75.Mn, 05.30.Jp, 67.85.Hj
\end{description}
\end{abstract}

\maketitle

\textit{Introduction.---}
Quantum vortex, as a type of topological defect, is an important topic in superfluids and superconductors (SC)~\cite{Wallraff2003, Abo-Shaeer2001, Blaauwgeers2000, Bugoslavsky2001}.
Some particular interesting quantum vortices include half-quantum vortex~\cite{PhysRevLett.108.010402, wu2011, hu2012}, vortex-antivortex pair~\cite{v-antiv}, fundamental vortex and so on.
In addition, a special attention has been payed to
vortex lattice, which determines the superconducting properties in type-II SC~\cite{klein2001}.
Quantum vortex and vortex lattice have been observed in variety of physical setting including Bose-Einstein condensates (BECs) with rotation~\cite{Abo-Shaeer2001}, Type-II SC~\cite{klein2001} and liquid helium~\cite{Blaauwgeers2000}.

Spin-orbit (SO) coupling has attracted many attentions both in condensed matter physics and ultracold atomic physics,
because it plays a center role in topological insulator, topological superconductor and quantum spin hall effect~\cite{RevModPhys.83.1057, RevModPhys.82.3045, Chen21122012}.
The experimental realization of BECs with one-dimensional~\cite{lin2011} and two-dimensional (2D)~\cite{2dbecsoc} SO coupling motivated a lot of researches both in theory~\cite{wu2011, benpra, benpre, wang2010, PhysRevLett.111.185303, PhysRevLett.111.185304, PhysRevLett.108.125301, PhysRevLett.108.010402, PhysRevLett.115.253902, PhysRevLett.107.200401, PhysRevA.86.033628, PhysRevLett.107.270401, PhysRevA.84.063624, PhysRevLett.110.235302, PhysRevLett.110.140407, PhysRevLett.109.015301, PhysRevA.91.013607, PhysRevLett.110.085304, zhang2012, cole2012, PhysRevLett.112.180403} and experiment~\cite{PhysRevLett.109.115301, PhysRevLett.114.070401, PhysRevLett.114.105301, zhangnc2015, 10.1038/nphys2905}.
Up to now, the SO coupled BECs in a harmonic trap and free space~\cite{benpre, PhysRevLett.115.253902} have been studied widely, based on the numerical simulation of Gross-Pitaevskii equation (GPE).

SO coupled BECs in optical lattices are also studied based on GPEs~\cite{benpra, cole2012, PhysRevLett.112.180403}.
Cole \textit{et al.} reported that the energy minimums have four-fold degenerate structure and appear in diagonal directions in momentum space for case of weak interactions~\cite{cole2012}.
It was showed that the ground state in the setting holds lattice version of plane wave and stripe phase when energy minimums appear in diagonal directions~\cite{benpra}. 
It was also predicted that the half-quantum vortex soliton can appear even in four-fold degenerate structure for a SO coupled BEC in square Zeeman lattices~\cite{PhysRevLett.112.180403}.
However, the effect of SO coupling on the structure of vortex lattice formed in the optical lattice has not been explored. 
Particularly, the energy minimums  may appear in non-diagonal directions and in the boundary of the first Brillouin zone (BZ) as a consequence of interplay of SO coupling and optical lattice, which may induce the formation of exotic vortex lattices.

In this work, we focus on the competing region between the SO coupling strength $\lambda$ and reduced wave vector of optical lattices $\mathbf{k}$, the strong competition leads to a complex structure of energy minimums in momentum space.
At first, we aim to clarify the effect of optical lattices on the energy minimum of Rashba SO coupled BECs, especially when $\lambda$ is close to the boundary of the first BZ ($k_{OL}$).
Then, we research the exotic phases appeared in the competing region based on the degenerate structures of energy minimums.
\begin{figure}[H]
\includegraphics[width=0.44\textwidth]{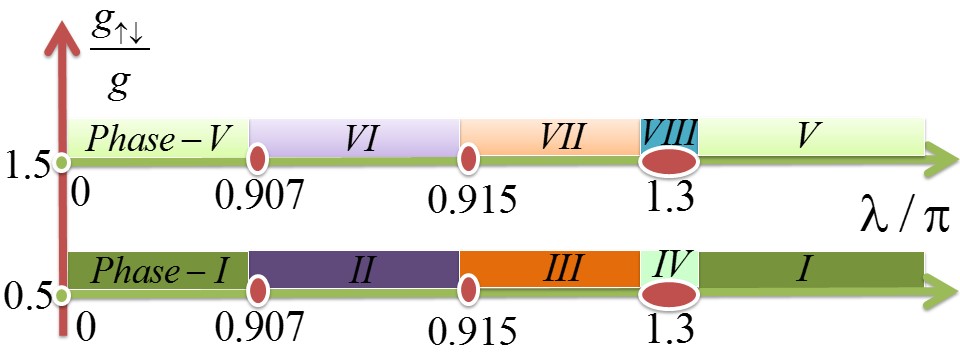}
\renewcommand{\figurename}{Fig}
\caption{(Color online)
Phase diagram of ground state of 2D SO coupled BECs in square optical lattices.
}
\label{phase-diagram}
\end{figure}
\begin{figure*}
\subfigure{\includegraphics[width=0.20\textwidth]{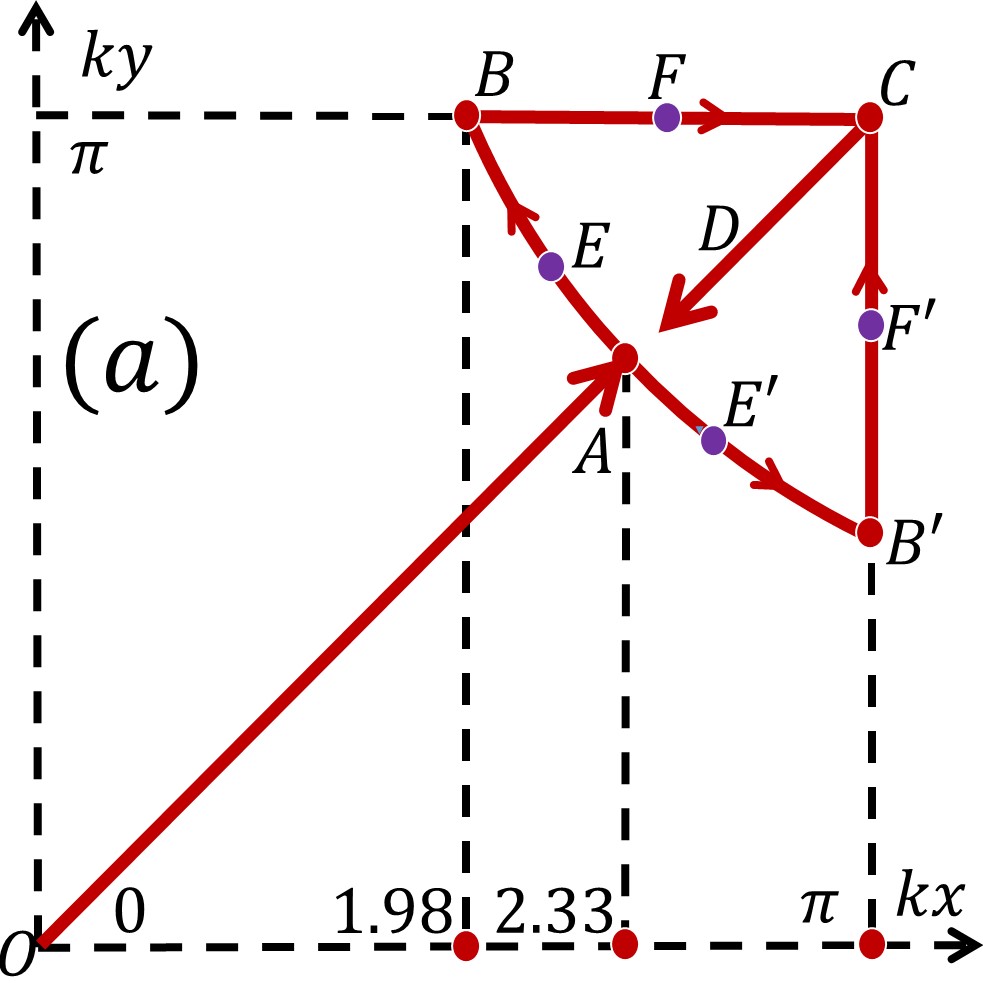}}
\subfigure{\includegraphics[width=0.78\textwidth]{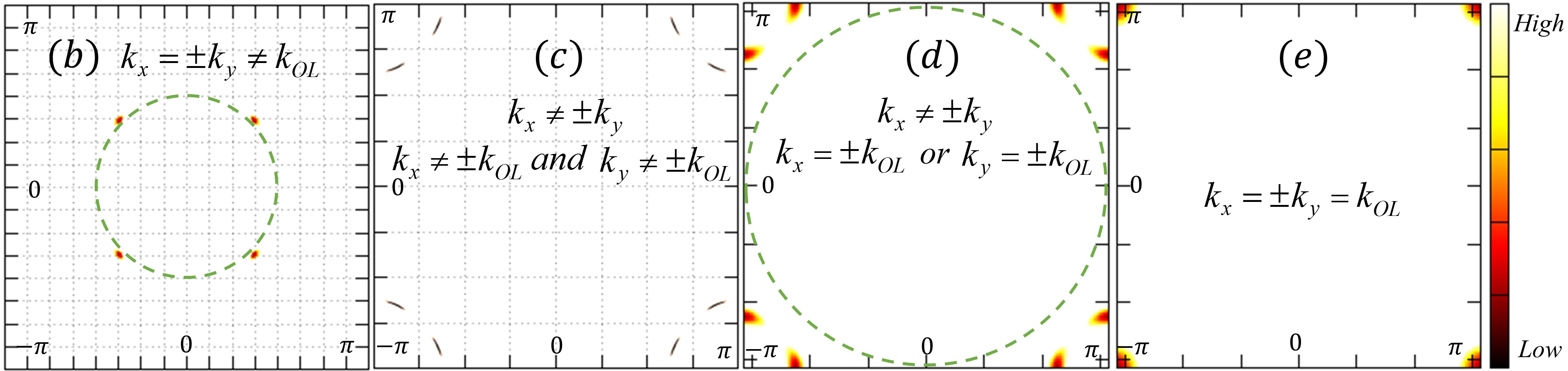}}
\renewcommand{\figurename}{Fig}
\caption{(Color online)
$(a)$ Evolution of energy minimums with increasing $\lambda$ in first quadrant of the first BZ.
$(b - e)$ Single particle energy minimums of ground state for 2D SO coupled BECs in optical lattices.
The figures from $(b)$ to $(e)$ correspond to $\lambda=\pi/2, 0.909\pi, \pi, 1.3\pi$.
$(c)$ The eight minimums are located in $(k_x = \pm 2.75, k_y = \pm 2.05)$ and $(k_x = \pm 2.05, k_y = \pm2.75)$.
$(d)$ The eight minimums are located in $(k_x = \pm 2.255, k_y = \pm\pi)$ and $(k_x = \pm\pi, k_y = \pm2.255)$.
$(e)$ The four minimums are located in $(k_x = \pm\pi, k_y = \pm\pi)$.
The green rings in $(b)$ and $(d)$ indicate the Rashba ring for SO coupled BECs without optical lattices.
}
\label{single-particle}
\end{figure*}
\begin{figure*}
\centering
\subfigure{\includegraphics[width=0.78\textwidth]{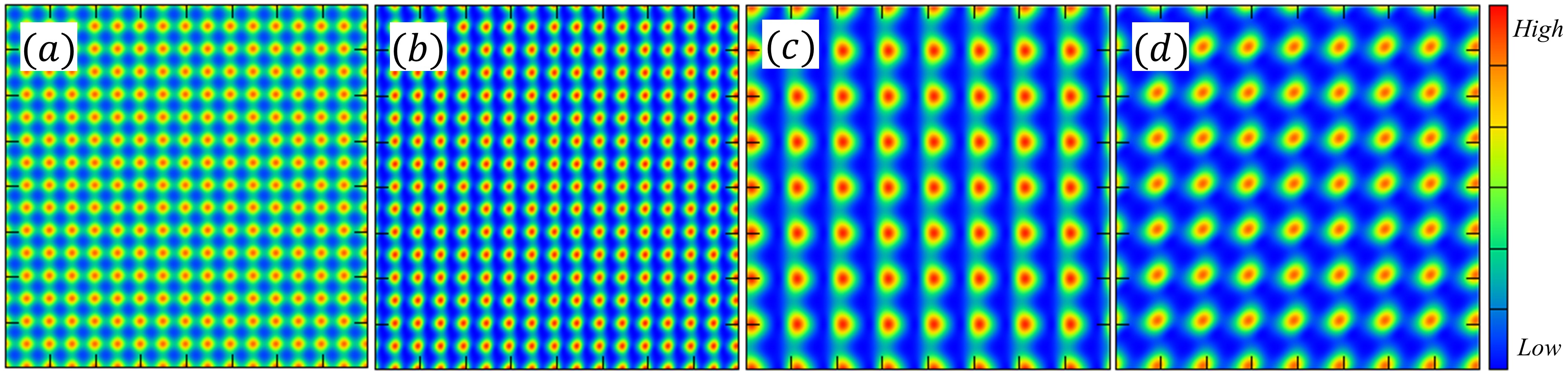}}
\subfigure{\includegraphics[width=0.78\textwidth]{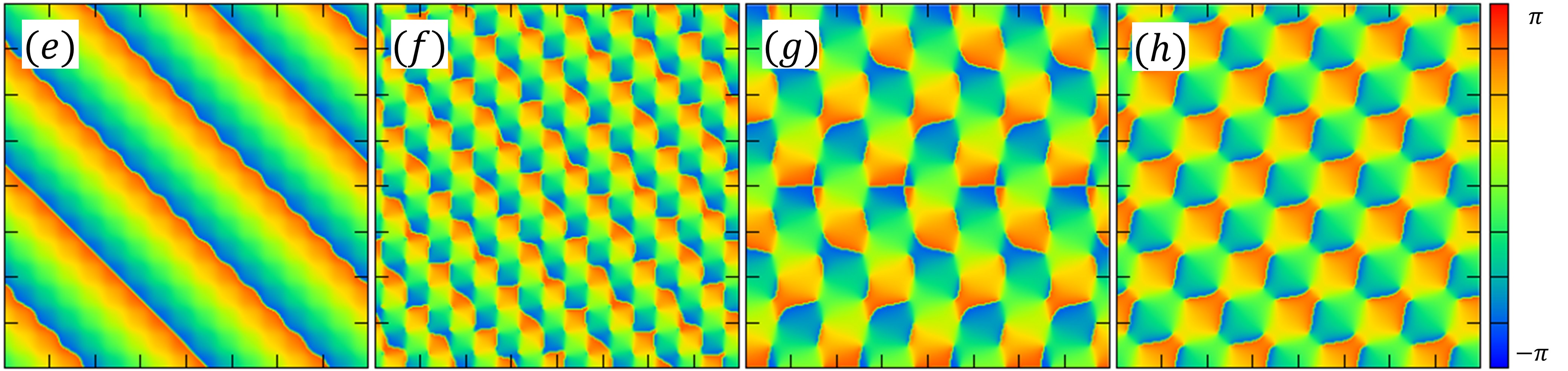}}
\renewcommand{\figurename}{Fig}
\caption{(Color online)
Density (upper row) and phase (lower row) distributions of spin-up components of phase I-IV.
From left to right, $\lambda=\pi/2, 0.909\pi, \pi, 1.3\pi$ and $N_x=N_y=16, 16, 8$ and $8$.
}
\label{dps}
\end{figure*}

\begin{figure}
\centering
\subfigure{\includegraphics[width=0.235\textwidth]{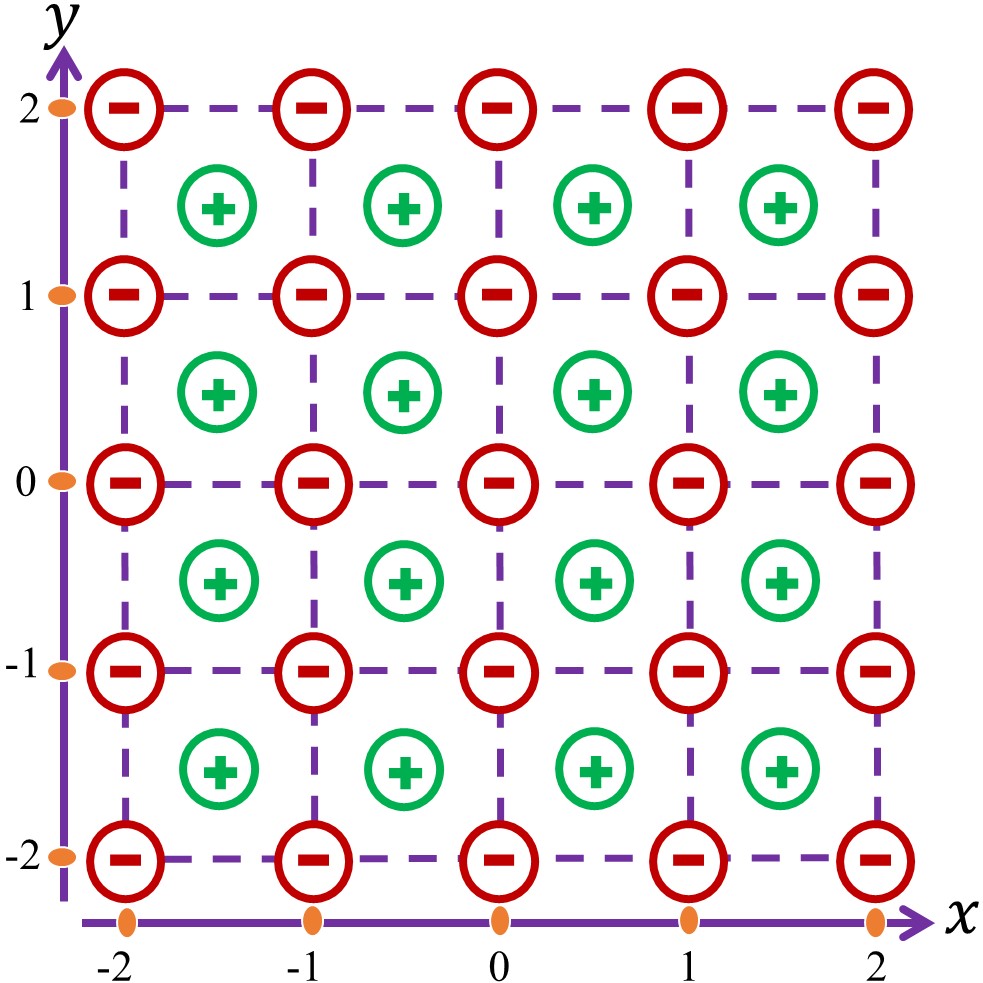}}
\subfigure{\includegraphics[width=0.235\textwidth]{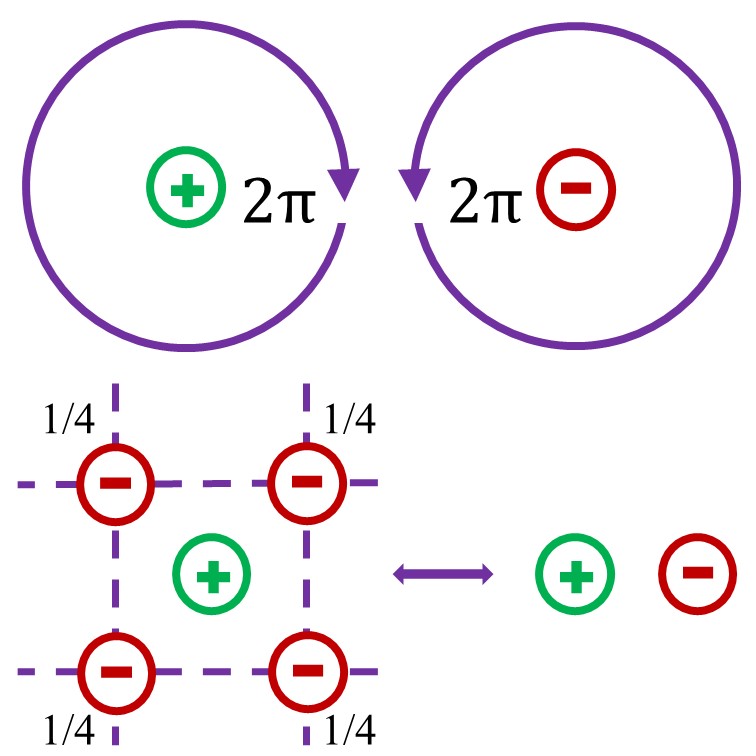}}
\renewcommand{\figurename}{Fig}
\caption{(Color online)
Multi-fold vortex lattice structures in spin-up component of phase-VIII.
}
\label{phase4}
\end{figure}

\begin{figure*}
\centering
\subfigure{\includegraphics[width=0.78\textwidth]{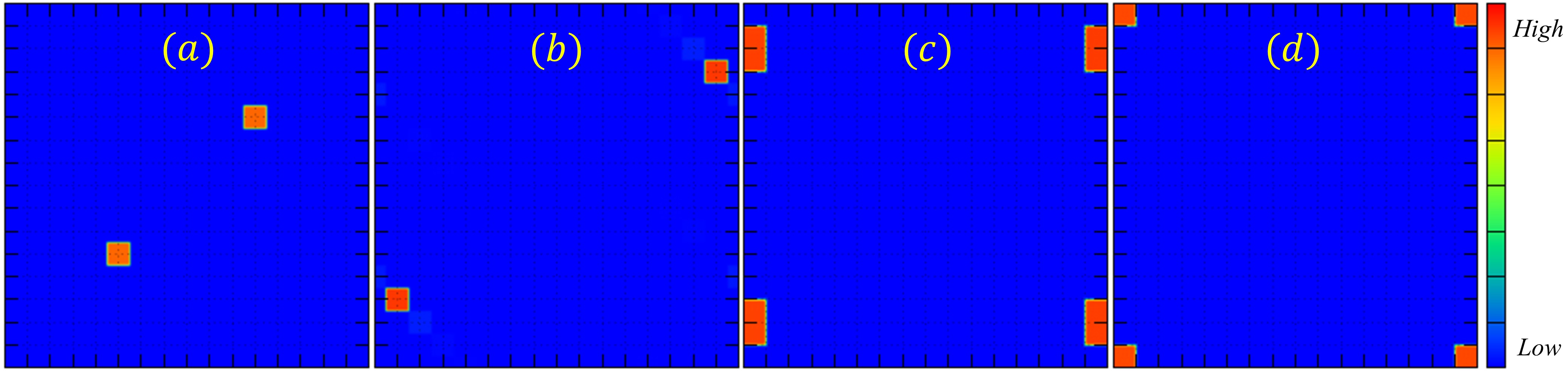}}
\subfigure{\includegraphics[width=0.78\textwidth]{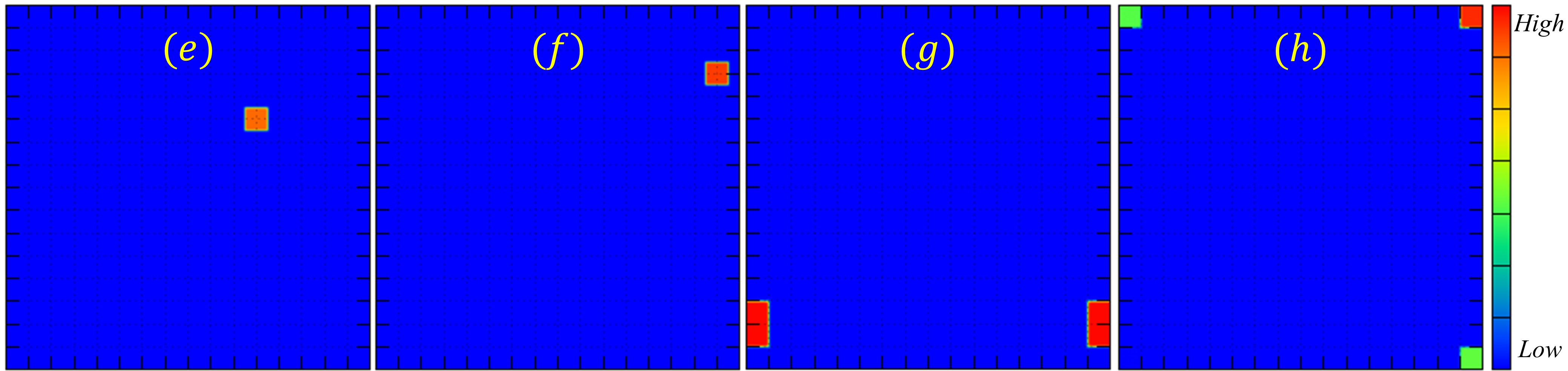}}
\renewcommand{\figurename}{Fig}
\caption{(Color online)
Momentum distributions of ground state for 2D SO coupled BECs in square optical lattices.
\textbf{Upper row}: phase V-VIII.
\textbf{Lower row}: phase I-IV.
The figures from left column to right column correspond to $\lambda=\pi/2, 0.909\pi, \pi, 1.3\pi$.
}
\label{ek}
\end{figure*}

The main results of this work are summarized, as a phase diagram of ground state, in Fig.~\ref{phase-diagram}. 
(I) With increasing $\lambda$, the system displays rich phases with different lattice structures, which correspond to the single particle spectrum having four-, eight-, pseudo-eight- and pseudo-four-fold degenerate structure, as shown in Fig.~\ref{single-particle}(a).
(II) The ground state holds two types of phases indicated by parity-time-reversal $(PT)$ (phase I-VII) or parity $(P)$ (phase-VIII) symmetry. 
Especially, phase-VIII with $P$ symmetry has multi-fold vortex lattice structures including half-quantum vortex lattices, vortex-antivortex pair lattices, and fundamental vortex lattices, as shown in Fig.~\ref{phase4}.
(III) Phase I-IV or phase V-VIII occupy the ground state, depending on the ratio between  inter-component interaction ($g_{\uparrow\downarrow}$) and intra-component interaction ($g$). 
The density and phase distributions of phase I-VIII can be understand from the momentum distributions of ground state, as shown in Fig.~\ref{ek}.
\begin{figure*}
\centering
\subfigure{\includegraphics[width=0.98\textwidth]{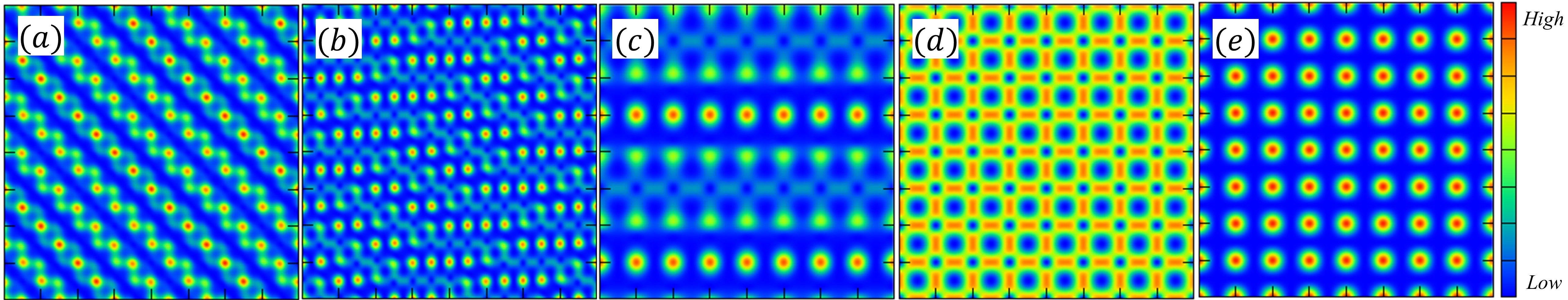}}
\subfigure{\includegraphics[width=0.98\textwidth]{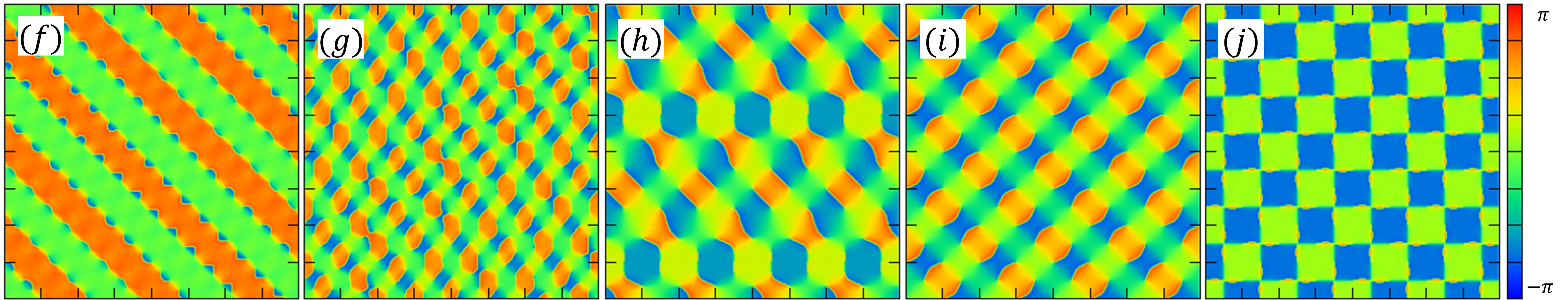}}
\renewcommand{\figurename}{Fig}
\caption{(Color online)
Density $(a - d)$ and phase $(f - i)$ distributions of spin-up components of phase V-VIII.
From left to right, $\lambda=\pi/2 (a, f), 0.909\pi (b, g), \pi (c, h)$ and $1.3\pi (d, i)$. 
For cases of $\lambda=\pi/2$ and $0.909\pi$, $N_x=N_y=16$, otherwise, $N_x=N_y=8$.
The right column corresponds to density $(e)$ and phase $(j)$ distributions of spin-down component of phase-VIII when $\lambda=1.3\pi$.
}
\label{dpl}
\end{figure*}

\textit{Model and Methods.---}
The Hamiltonian of Rashba SO coupled BECs in square optical lattices can be written as $\hat{H}=\hat{H}_0+\hat{H}_{int}$, here the single particle Hamiltonian
\begin{equation}
\hat{H}_0=\int d\mathbf{r}\mathbf{\psi}^\dag\left[-\hbar^2\nabla^2/(2M)+V_{OL}+h_{SO}\right]\psi,\label{subeq:2}
\end{equation}
where $\psi=(\psi_{\uparrow}, \psi_{\downarrow})$ is the wave function of spin-$1/2$ system, and $V_{OL}=-V_0 [\cos(2\pi x/d)+\cos(2\pi y/d)]$ describes the optical lattice with $V_0$ being the strength of optical lattice and $d$ being the period of lattice. 
Without loss of generality, we fix $V_0=5$ and $d=1$ throughout this work.
The SO coupling term is given by $h_{SO}=\lambda(k_x\sigma_x+k_y\sigma_y)$, where $\sigma_x$ and $\sigma_y$ express the Pauli matrices. 
The interaction term
$\hat{H}_{int}=\int d\mathbf{r}\left[g\hat{n}_{\uparrow}
^2+g\hat{n}_{\downarrow}^2+2g_{\uparrow\downarrow}\hat{n}_{\uparrow}\hat{n}_{\downarrow}\right]$,
where $\hat{n}_{i}=\mathbf{\psi}^\dag_{i}\mathbf{\psi}_{i}$ and $i=\uparrow$ or $\downarrow$ expresses spin-up or spin-down component, respectively.

The wave function of the single particle Hamiltonian (Eq.~\ref{subeq:2}) can be constructed by Bloch state and spin wave function.
The Bloch state for each component of spinor condensate has a form of $\psi_{i}(\mathbf{r})=e^{i\mathbf{k}\cdot\mathbf{r}}\phi_{i}(\mathbf{r})$, here $\phi_{i}(\mathbf{r})$ is a periodic function with period of optical lattices.
The reduced wave vector $\mathbf{k}_{i}=2\pi l_{i}/N_i d$ $(i=x, y)$, here $l_i$ is an integer, $N_i$ indicates the number of lattice cell in $i$-direction, and $\mathbf{k}_{i}\in[-\pi/d,\pi/d]=[-k_{OL},k_{OL}]$.

\textit{Single particle spectrum.---}
The single particle spectrum can be obtained by numerically solving Eq.~(\ref{subeq:2})~\cite{benpra} as shown in Fig.~\ref{single-particle}.
In Fig.~\ref{single-particle}(a), we show the evolution of energy minimums in first BZ as a function of $\lambda$.
There are four types of degenerate structure of energy minimums:
(I) four-fold degeneracy (when $0<\lambda<\lambda_{c1}=0.907\pi$ or $\lambda >\lambda_{c3}=1.3\pi$),
(II) eight-fold degeneracy (when $\lambda_{c1}\leqslant\lambda<\lambda_{c2}=0.915\pi$),
(III) pseudo-eight-fold degeneracy (when $\lambda_{c2}\leqslant\lambda<\lambda_{c3}$),
and (IV) pseudo-four-fold degeneracy (when $\lambda=\lambda_{c3}$).
For BECs without SO coupling in optical lattices, the energy minimum occurs in the origin of BZ.
As increasing $\lambda$, the energy minimums undergo the four-fold degeneracy along $\overrightarrow{OA}$ until $\lambda$ approaches to $\lambda_{c1}$.
Then, increasing $\lambda$ until $\lambda_{c2}$, the degenerate structure changes to eight-fold degeneracy, i.e., the energy minimum splits from one-point $A$ to two-points $E$ and $E^{\prime}$ in $\overrightarrow{AB}$ and $\overrightarrow{AB^{\prime}}$.
Next, the energy minimums fall into structure-III, where the energy minimums close to $C$ from two points $B$ and $B^{\prime}$ along $\overrightarrow{BC}$ and $\overrightarrow{B^{\prime}C}$, and the minimums appear in $k_{x} = \pi$ and $k_{y} = \pi$ directions, respectively.
However, due to the periodicity of BZ, the minimums with $k_{x, y}=\pi$ and $k_{x, y}=-\pi$ are the same one, therefore, the structure-III is indeed four-fold degeneracy.
Furthermore, when $\lambda=\lambda_{c3} = 1.3\pi$, the energy minimums appear in the point C with four degenerate minimums in the corners of BZ being the same due to the periodicity of BZ.
Finally, when $\lambda>\lambda_{c3} = 1.3\pi$, the energy minimums go back to the four-fold degenerate structure along $\overrightarrow{CD}$.
For case of four-fold degeneracy, the minimums occur in directions $k_y = \pm k_x$, in agreement with Refs.~\citep{benpra, cole2012}.
In Fig.~\ref{single-particle} $(b-e)$, we show the energy minimums in fist BZ for $\lambda=\pi/2$, $0.909\pi$, $\pi$ and $1.3\pi$ corresponding to degenerate structure I-IV, respectively.

We want to point out that: 
(I) The radius of Rashba ring for SO coupled BECs in optical lattices is not equal to $\lambda$ exactly, because it is modified by optical lattices.
When $\lambda$ is significantly smaller than $k_{OL}$ (Fig.~\ref{single-particle}$(b)$), the minimums are almost taken from Rashba ring with a slight offset.
On the other hand, for case of $\lambda \sim k_{OL}$ (Fig.~\ref{single-particle}$(d)$), the degenerate minimums are far away from Rashba ring due to strong competition of $\lambda$ and $k_{OL}$.
(II) To distinguish the adjacent degenerate minimums in first BZ, we need to select a suitable $N_i$, for our case, we take $N_i=16$ or $8$ for different degenerate structures.

\textit{Ground state phase diagram.---}
We address the ground state of SO coupled BECs in optical lattices by solving the GPEs in the framework of mean-field theory~\cite{benpra,benpre}.
The interactions ($g$ and $g_{\uparrow\downarrow}$) take a role of breaking the degeneracy of single particle states, and make the ground state holds more rich phases.
We found that the ground state includes two-types of phases with $PT$ or $P$ symmetry, respectively.
Among them, phase (I-VII) possess $PT$ symmetry with $n_{\uparrow}(\mathbf{r}) = n_{\downarrow}(\mathbf{-r})$~\cite{PhysRevLett.108.010402}, here $n_{i}(\mathbf{r})$ indicates the density of condensates in space $\mathbf{r}$.
On the other hand, phase-VIII possesses $P$ symmetry with $n_{i}(\mathbf{r}) = n_{i}(\mathbf{-r})$.

For case of $g>g_{\uparrow\downarrow}$, the ground state falls into phase-(I-IV), as shown in Fig.~\ref{dps}.
Due to $g$ dominates the system, spin-up and spin-down components occupy same lattice cell, meanwhile $g_{\uparrow\downarrow}$ makes two components located in two sides of the center of each lattice cell in $x$ and $y$ directions or both.
As a result, the condensates form ordered lattices, i.e., every lattice cell has same density distribution.
As expected, phase-I is a lattice version of plane wave phase appeared in homogeneous case~\citep{wang2010}.
Phases (II-IV) have similar density distribution with phase-I, but different phase distribution, 
because of either or both of $\mathbf{k}_x \neq\mathbf{k}_y$ in momentum space and the periodicity of BZ.

On the other hand, for case of $g<g_{\uparrow\downarrow}$, phases (V-VIII) occupy the ground state, the density and phase distributions are shown in Fig.~\ref{dpl}. 
Comparing to phases I-IV,
the stronger $g_{\uparrow\downarrow}$ makes different spin components of phases V-VII occupy alternatively the adjacent lattice cells.
Phase-V with a stripe structure consists of several chains in both components, and each chain includes several plane waves.
In particular, we want to point out that the area with phase $-\pi$ in phase distribution is decreasing as increasing $g$.
When the interaction energy is significantly larger than lattice potential energy, the lattice stripe phase falls into stripe phase appeared in harmonic trap case~\citep{PhysRevLett.107.270401} and homogeneous case~\citep{wang2010}.
The density distribution of phase-VI has a complexed pattern induced by the mismatch between the direction of the primitive translation vectors of optical lattices and the stripe direction, as shown in Fig.~\ref{dpl}$(b)$.
The diagonal or non-diagonal stripe structure for phase-(V or VI) can be understand from $\mathbf{k}_x=\mathbf{k}_y$ or $\mathbf{k}_x \neq \mathbf{k}_y$ in momentum distribution (Fig.~\ref{ek} $(b)$).
Phase-VII shows a stripe distribution in $y$ direction and coexistence of soliton-like wave and vortex in both components.
Combining two components together, we found that Phase-VII holds two sets of half-quantum vortex lattices.
The first set is consisting of vortices in spin-up component and soliton-like wave in spin-down component, and another set is consisting of soliton-like wave in spin-up component and vortices in spin-down component.

\textit{Multi-fold vortex lattices.---}
As an exotic phase, phase-VIII with $P$ symmetry holds multi-fold vortex lattices (Fig.~\ref{phase4}) induced by the strongest competition of SO coupling and optical lattices.
The density and phase distributions are shown in Fig.~\ref{dpl}$(d, e, i, j)$, the stronger $g_{\uparrow\downarrow}$ makes two components separate to crest and trough of optical lattices, respectively. 
The first vortex lattices appeared in phase-VIII is half-quantum vortex lattices in which each lattice cell carries a half-quantum vortex, i.e., spin-up component holds an anti-vortex (indicated by $''-''$) while spin-down component holds a soliton-type wave, so that, in whole lattices, the condensates form a half-quantum vortex lattices. 
The second one is vortex-antivortex~\cite{v-antiv} pair (indicated by $''+''$ and $''-''$) lattices in spin-up component induced by optical lattices and SO coupling, where vortex appear in crest of optical lattices around by four anti-vortices located in trough of optical lattices.
The last one is fundamental vortex lattices (indicated by $''+''$) in the spin-up component induced by optical lattices.
Phase-VIII can be regarded as a lattice counterpart of half-quantum vortex state appeared in case of harmonic trap~\citep{hu2012, wu2011} or free space~\citep{benpre}, but the presence of optical lattice makes the phase holds exotic  vortex lattices structure.

\textit{Momentum distribution of ground state.---}
The momentum distribution of ground state is shown in Fig.~\ref{ek}.
Here, the wave function in momentum space can be written as $\psi_{i}(\mathbf{k})=\int d\mathbf{r}e^{-i\mathbf{k}\cdot\mathbf{r}}\psi_{i}(\mathbf{r})$ $(i=\uparrow,\downarrow)$.
Comparing to single particle results (Fig.~\ref{single-particle}$(b-e)$), we confirmed that the many-body ground state is formed by selecting some minimums from single particle spectrum.    

In Fig.~\ref{ek} $(a,b,e,f)$, we show the momentum distributions when $\lambda=\pi/2$ and $0.909\pi$.
For cases of $g<g_{\uparrow\downarrow}$, two minimums with opposite momentum in single particle case are selected to form phase-(V, VI) holding two- and four-fold degeneracy, respectively.
On the other hand, for case of $g>g_{\uparrow\downarrow}$, one minimum in single particle case are selected to form phase-(I, II) holding four- and eitht-fold degeneracy, respectively.
The difference in density and phase distribution between phase-I(V) and phase-II(VI) is originated from $\mathbf{k}_x=\pm \mathbf{k}_y$ and $\mathbf{k}_x \neq \pm \mathbf{k}_y$ in momentum distribution, respectively. 

The momentum distributions for cases of $\lambda=\pi$ and $1.3\pi$ reflect the periodicity of BZ, as shown in Fig.~\ref{ek}$(c,d,g,h)$.
For case of $\lambda=\pi$, the single particle spectrum has pseudo-eight-fold degeneracy, which is actually four-fold degeneracy due to the equivalence of $k_{x, y}=\pi$ and $-\pi$ in first BZ.
Therefore, the momentum distribution for phase-VII with two-fold degeneracy can be regarded as two minimums located in upper-right and lower-left (see Fig.~\ref{ek}$(c)$) when $g<g_{\uparrow\downarrow}$.
On the contrary, the momentum distribution for phase-III with four-fold degeneracy can be regarded as one minimum located in lower-left or lower-left (see Fig.~\ref{ek}$(g)$) when $g>g_{\uparrow\downarrow}$.
For case of $\lambda=1.3\pi$, the momentum distribution with minimums in $\mathbf{k}_{x, y}=\pm\pi$ holds the highest symmetry.
Due to the periodicity of BZ, all of the minimums in single particle spectrum are equivalent.
For case of $g<g_{\uparrow\downarrow}$, all of the minimums are selected to form the phase-VIII with signature of multi-fold vortex lattice (see Fig.~\ref{ek}$(d)$).
On the other hand, for case of $g>g_{\uparrow\downarrow}$, three minimums are selected to form the phase-IV, but two minimums of them in diagonal direction just contribute half of their density indicated by green points in Fig.~\ref{ek}$(h)$.
This is quite different from other cases, in which all minimums contribute full density to form a phase once the minimums are selected.

\textit{Conclusions.---}
We have investigated the energy spectrum and ground states of 2D Rashba SO coupled BECs in square optical lattices in competing region of $\lambda$ and $\mathbf{k}$.
When $\lambda \sim k_{OL}$, we found that the energy spectrum holds eight-, pseudo-eight- and pseudo-four-fold degenerates structure due to periodicity of BZ.
Furthermore, we showed that the momentum distributions of interacting BECs are formed by selecting some energy minimums in single particle case.
On the other hand, the density and phase distributions of ground state have important and fascinating patterns.
Especially, we found that multi-fold vortex lattice structure, in which half-quantum vortex lattice, vortex-antivortex pair lattice and fundamental vortex lattices co-exist, occupies the ground state when $\lambda = 1.3\pi$.
Finally, the ground state can be divided to two types with $PT$ or $P$ symmetry.

\textit{Acknowledgments.---}
The work is supported by NSFC under Grants No. 11425419, No. 11374354 and No. 11174360, MOST project under the contract No. 2016YFA0300603 and the Strategic Priority Research Program (B) of the Chinese Academy of Sciences (No. XDB07020000).

\bibliography{SOC-BEC-Bibtex-20160918-jpeg}
\end{document}